\documentstyle{article}
\begin{document}
\title{\bf Imprints of Nonextensivity in Multiparticle 
Production\footnote{Invited talk presented by G.Wilk
at {\it $6^{th}$ International Workshop on Relativistic Aspects
of Nuclear Physics} (RANP2000), Caraguatatuba, Tabatinga Beach,
S\~ao Paulo, Brazil, October 17-20, 2000.} }
\author{G.Wilk$^{1}$\thanks{e-mail: wilk@fuw.edu.pl} and Z.W\l
odarczyk$^{2}$ \thanks{e-mail: wlod@pu.kielce.pl}\\[2ex] 
$^1${\it The Andrzej So\l tan Institute for Nuclear Studies}\\
    {\it Ho\.za 69; 00-689 Warsaw, Poland}\\
$^2${\it Institute of Physics, Pedagogical University}\\
    {\it  Konopnickiej 15; 25-405 Kielce, Poland}}  
\date{\today}
\maketitle

\begin{abstract}
The statistical methods based on the classical Boltzmann-Gibbs (BG)
approach are at heart of essentially all descriptions of
multiparticle production processes. In many cases, however, one
observes some deviations from the expected behaviour. It is also
known that conditions necessary for the BG statistics to apply are
usually satisfied only approximately. Two attitudes are possible in
such situations: either to abandon statistical approach trying some
other model or to generalise it to the so called nonextensive
statistics (widely used in the similar circumstances in many other
branches of physics). We shall provide here an overview of possible
imprints of non-extensitivity existing both in high energy cosmic ray
physics and in multiparticle production processes in hadronic
collisions, in particular in heavy ion collisions.\\

[3ex]

\end{abstract}

\section{Introduction}

The high energy collisions are usually connected with production of
large number of secondaries (mostly $\pi$ and $K$ mesons). The strong
interactions involved here make their detail descriprion {\it from
first principles} impossible and one is forced to turn to
phenomenological models of various kinds. The statistical models were
the first successful approaches to the multiparticle production
processes since the beginning of the subject almost half century ago
\cite{LANDAU} and they remain still very much alive today, especially
in all analysis of multiparticle data performed from the point of
view of the possible formation of the new state of matter - the Quark
Gluon Plasma (QGP) \cite{QGP}. They are all based on the
Boltzmann-Gibbs form of entropy, which is identical to the so called
Shannon entropy used in the information theory approach.\\ 

We would like to stress at this point that information theory can be
also applied to hadronic processes \cite{MAXENT}. Its importance
there is best visualized by the following example. Suppose that some
experiment provides us with data on a limited set of $n$ observables:
$R_1, R_2, \dots , R_n$. This triggers the theory, which follows with
a variety of models, usually completely different (if not
contradictory) but each claiming good agreement with these data. The
models provide therefore apparently different views of the same
physical reality (data) leading to confusion on what is actually
going on. It happens this way because experimental data contain only
{\it limited amount of information} to which theoretical models add
their own specific assumptions born not in data but in our minds. To
quantify this situation one must find the way to answer (in a model
independent way) the following question: given physical assumptions
of a model (plus phase space and conservation laws), what are their
{\it most trivial} consequences? This can be done only be by
resorting to information theory which defines {\it triviality} as the
{\it lack of information} (the less information the more trivial).
The measure of information provided by Shannon entropy (or some other 
kind of entropy, in our talk this will be Tsallis entropy \cite{T}) 
allows us then to quantify the whole problem.

Let us elaborate this point a bit further. Experiment gives us
usually set of points to which one tries to fit some distribution
function $\rho(x)$, which can be thought to be some probability
distribution. In this case Shannon entropy corresponding to it is
\begin{equation}
S\, =\, - \int \rho(x)\, \ln[\rho(x)] . \label{eq:Shannon}
\end{equation}
We are looking for the most trivial $\rho(x)$, i.e., for one
containing the least information, and it means for such, which
informational entropy is maximal under given circumstances
(constraints)\footnote{One should remember that formula
(\ref{eq:Shannon}) assumes that $\rho(x)$ is measured against some
other distribution $\rho_0(x)$ corresponding to equal {\it a priori}
probability being assigned to equal volumes of phase space. This is
sometimes called "the first principle of statistical mechanics"
\cite{MAXENT}. The Boltzmann constant, which normally appears in
(\ref{eq:Shannon}) has been henceforth assumed unity, $k_B =1$.}. 
What are those constraints? At first, being a probability distribution
$\rho(x)$ must be non-negative and normalized, i.e.,
\begin{equation}
\int \rho(x)\, dx \, =\, 1 . \label{eq:norm}
\end{equation}
Further, $\rho(x)$ has to reproduce the experimental results given in
the form of expectation values $\langle R^{(k)(x)}\rangle = R_k$ of
measured set of observables $R^{(k)}$, i.e.,
\begin{equation}
\int \, dx\, R^{(k)}(x)\, \rho(x)\, =\, R_k ;~k=1, \dots ,m . 
                        \label{eq:R}
\end{equation}
Using eqs.(\ref{eq:norm}) and (\ref{eq:R}) as constraints one then
maximalizes information entropy (\ref{eq:Shannon}) and as results
one gets the seek for distribution $\rho(x)$,
\begin{equation}
\rho(x)\, =\, \frac{1}{Z}\cdot \exp\left[ - \sum_{k=1}^m\, \beta_k\,
               R^{(k)}(x) \right],  \label{eq:FINDIS}
\end{equation}
where $Z$ is calculated from the normalization condition. This
distribution reproduces known information (i.e., it tells us {\it the
truth, the whole truth} about experiment) and contains the least
information (i.e., it tells us {\it nothing but the truth}). In this
sense it is the unique, most plausible (least biased), model
independent probability distribution describing the outcomes of our 
experiment. The Lagrange multipliers $\beta_k$ are given as solutions
of the constraint equations and $R_k = \partial \ln
Z(\beta_1,\dots,\beta_m)/\partial \beta_k$. \\

Now we can understand why apparently completely disparate models
can describe sucessfully the same experiment. It is because they share
(usually in an implicite than explicite way) the same set of basic
common assumptions (i.e., constraints $R_k$) and, at the same time,
they differ completely in the rest of them, i.e., in those $R_k$,
which contain assumptions particular for a given model (and which are
completely not relevant to its agreement with experiment)\footnote{A
completely forgotten example 
of such situation is discovery \cite{Chao} that to explain all
experimental data on the multiparticle reactions of that time it is
only necessary to assume that the phase space is effectively
one-dimensional (i.e., the $p_T$ cut-off) and that not the whole
initial energy $\sqrt{s}$ of reaction is used for the particle
production (i.e., the existence of the inelasticity $K$ or of the
leading particle effect).}.\\ 

This example tells us that statistical approach is not necessarily
connected with any thermodynamical model but has, in fact, much
broader range of applications. In our example above the formula
(\ref{eq:FINDIS}) is identical with formulas used in thermal 
models \cite{QGP} but parameters $\beta_k$ can take any values
depending on the energy and multiplicity of the event under
considerations \cite{MAXENT}. It is worth to remind at this point
the very old observation \cite{EXP} that exponential form
(\ref{eq:FINDIS}) of so many distributions observed experimentally
can have its origin in the fact that the actual measurements extend
only to subsystems of a total system. The summation over the
unmeasured (i.e., essentially overaged over) degrees of freedom
introduces the randomnization, which can be most economically
described as a kind of thermal bath leading to thermal-like
exponential distributions with some effective "temperature "$T$. \\

In this approach dynamics lies in our ability to reproduce the known
information (i.e., is hidden in $\beta_k$). However, the big
advantage of this method is that it prevents us from smuggling in
anything additional, not being present in data. The outcome depends
on the funtional form of the entropy. The Shannon form
(\ref{eq:Shannon}) is the most popular one because it coincides in
thermodynamical limit with the Boltzmann-Gibbs (BG) form allowing for
contact with standard thermodynamics. However, as newtonian mechanics
is not eternal and universal (in the sense that in the limit of large
velocities it must be replaced by special relativity, in the limit of
vanishingly small masses quantum mechanical effects must be accounted
for and in the limit of extremaly large masses (stars, galaxies,...)
it must be replaced by general relativity) the same can be said about
the classical BG statistical mechanics. It is fully satisfactory only
if
\begin{itemize}
\item effective microscopic interactions are short-ranged (i.e., we
have close spacial connections);
\item effective microscopic memory is short-ranged (i.e, we have close
time connections);
\item boundary conditions are non (multi)fractal (i.e., the relevant
space-time and/or phase space is non (multi)fractal).
\end{itemize}
As a matter of fact despite the popularity of BG it seems that in
majority of situations we are dealing with, at least one of these
conditions is not satisfied. Therefore it is justified to look for
the possible generalization of BG which would provide a
compactifications of all possible deviations from the ideal BG
statistics into a minimal - possible only one - additional
parameters. In this way we arrive at the notion of Tsallis statistics
based on Tsallis entropy \cite{T}:
\begin{equation}
S_q\, =\, -\, \frac{1 - \int dx \rho(x)^q}{1 - q} .
\end{equation}
We shall not dwell too much on it here as this is the subject of
separate lecture by Prof. Tsallis himself \cite{TT}. The only points 
we would like to stress here for our purposes are:
\begin{itemize}
\item for $q \rightarrow 1$ one recovers the usual Shannon entropy
form of (\ref{eq:Shannon});
\item with the normalization condition imposed on $\rho(x)$ as before
and with the new definition of the mean values, 
$\langle R\rangle_q = \int dx R(x) \rho(x)^q $,
one replaces formula (\ref{eq:FINDIS}) with
\begin{equation}
\rho(x)\, =\, \frac{1}{Z}\cdot \left[ 1 - (1 - q)\sum^m_{k=1} \beta_k
R^{(k)}(x)\right]^{\frac{1}{1-q}} ; \label{eq:FINDISq}
\end{equation}
(notice that, again, for $q\rightarrow 1$ (\ref{eq:FINDISq}) goes to
(\ref{eq:FINDIS}). As before, this distribution reproduces known
information and conveys least information but this time information
is measured in a different way, which is paramatrized by the
parameter $q$)\footnote{There is a formalism, which expresses both
the Tsallis entropy and the expectation value using the escort
probability distribution, $P_i = p_i^q/\sum p_i^q$\cite{T}. However,
it is known \cite{Abe} that such an approach is different from the
present nonextensive formalism since the Tsallis entropy expressed in
terms of the escort probability distribution has a difficulty with
the property of concavity.}; 
\item the resultant staistics is non-extensive in the sense that 
for two subsets $A$ and $B$ such that probabilities
$p_{ij}(A+B)=p_i(A)p_k(B)$ the entropy is not additive but instead
\begin{equation}
S_q(A+B)\, =\, S_q(A)\,+\, S_q(B)\, +\, (1-q) S_q(A)S_q(B) \label{eq:Nonext}
\end{equation}
and additivity is recovered only in the limit $q\rightarrow 1$;
\item the whole Legendre structure of thermodynamics is preserved for
$q \neq 1$ \cite{T,TT}.
\end{itemize}

Although the nonextensive statistics were used for a long time in
many branches of physics \cite{T}, their possible imprints in
multiparticle production are very fresh. In this talk we shall
concentrate on some cosmic ray experiments and accelerator
experiments, which provided us with examples of a nonextensive
behaviour. Both will demonstrate traces of nonextensivity seen as
nonexponentiality of some distributions, i.e., where $\exp(x)
\Rightarrow \exp_q(q) = [1-(1-q)x]^{1/(1-q)}$. At the end we shall,
however, list all other works in which one can find similar
conclusions \cite{DENTON}.\\ 

\section{Cosmic Ray Example}

Part of the cosmic ray detector system consists of extrathick lead
emulsion chambers. Such chamber is build with a number of sandwiches
of X-ray sensitive film, plate of lead and layer of emulsion (first
allows to localize the point at which external particle of cosmic
rays enters, second forces it to interact and to start the
electromagnetic cascade process, and third allows to calculate
deposited energy). In some experiments one observes that distribution
of the points of first interaction (cascade starting points) does not
follow the expected simple exponential rule\footnote{Here $N$ is the
number of cascades originating at depth $T$ expressed in the so
called cascade units where $1$ c.u. = $6.4$ g/cm$^2$ = $0.56$cm of
Pb and $\lambda$ is the mean free path for the p-Pb
interacions.}\cite{WWCR}: 
\begin{equation}
\frac{dN}{dT}\, =\, {\rm const}\cdot \exp \left(\, -\,
\frac{T}{\lambda}\, \right) \label{eq:LFC}
\end{equation}
but deviates from it noticeably for larger $T$. This phenomenon
acquired even its own name: {\it long flying component} (because
original hadrons tend to fly longer without interaction than naively
expected).\\
\begin{figure}[h]
\setlength{\unitlength}{1cm}
\begin{picture}(25.,16.5)
\includegraphics{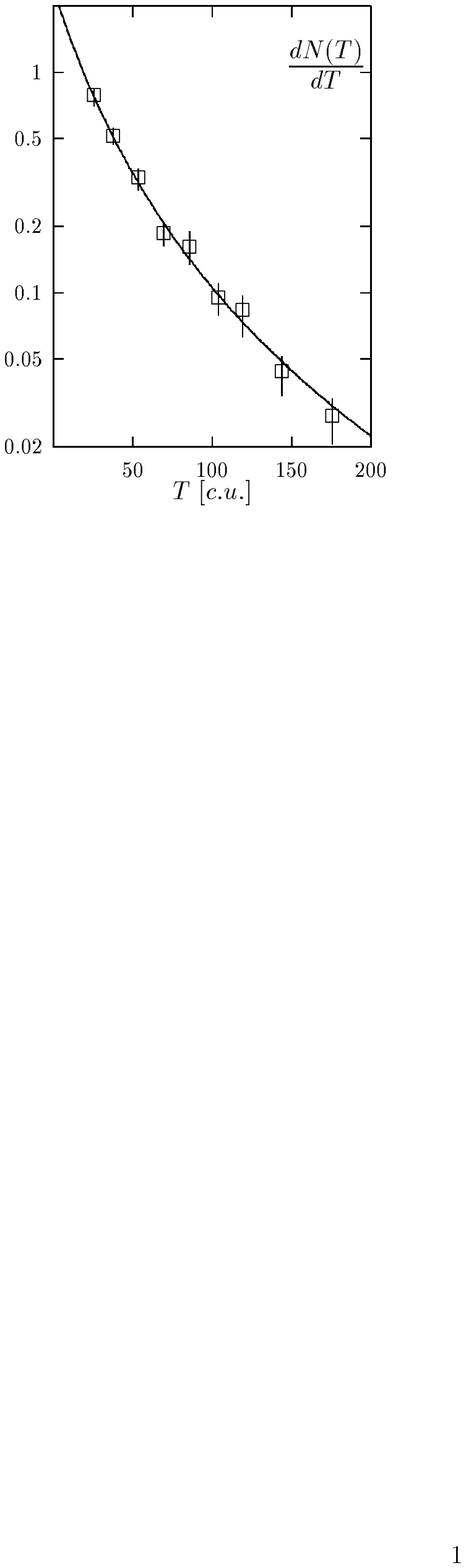}
\end{picture}
\end{figure}
\begin{itemize}
\item[Fig. 1] Depth distribution of the starting points,
              $dN(T)/dT$, of cascades in 
              Pamir lead chamber. Notice the non-exponential
              behaviour of data points (for their origin cf.
              \cite{WWCR}) which can be fitted by Tsallis
              distribution (\ref{eq:LFCQ}) with $q=1.3$. (This 
              figure is reproduced from Fig. 1 of \cite{CR}).
\end{itemize}

In \cite{WWCR} we have argued that the observed effect can be just
another manifestation of the fluctuation of the corresponding
hadronic cross section $\sigma = Am_N\frac{1}{\lambda}$ (where $A$
denotes the mass number of the target and $m_N$ is the mass of the
nucleon, such a possibility is widely discussed in the literature and
observed in diffraction dissociation experiments on accelerators, cf.
\cite{WWCR} for details and references). As was shown in \cite{WWCR}
such fluctuations of cross section (i.e., in effect $1/\lambda$) with
relative variance 
\begin{equation}
\omega\, =\, \frac{\langle \sigma^2\rangle - \langle \sigma \rangle ^2}
          {\langle \sigma \rangle ^2}\, \geq 0.2 \label{eq:omega}
\end{equation}
allow to describe the observed effect. However, it turns out
\cite{CR} that the same data can be fitted by the nonextensive
formula  
\begin{equation}
\frac{dN}{dT}\, =\, {\rm const}\cdot \left[1\, -\,
      (1-q)\frac{T}{\lambda}\right]^{\frac{1}{1-q}} \label{eq:LFCQ} 
\end{equation}
with parameter $q=1.3$ (in both cases $\lambda = 18.85\pm 0.66$ in
$c.u.$ defined above), cf. Fig. 1.\\

This example provides us hint on the possible physical meaning of the
nonextensvity parameter $q$. Comparing both explanations one can
suspect that perhaps, at least in this case, it has something to do
with the fluctuations of the seemingly constant parameter $1/\lambda$
in the exponential formula (\ref{eq:LFC}). We shall argue later that
this is indeed the case but before let us demonstrate another example
of this kind, now from the high energy multiparticle production
processes domain.

\section{Heavy ion collision example}

Heavy ion collisions are of particular interest at present because
they are the only place were one expects the possible formation of
the QGP - a new state of matter \cite{QGP}. The termal and/or all
kind of statistical models are playing very important role here
because one of the crucial parameter is the temperature $T$ of
reaction deduced usualy from the transverse momentum distributions,
$dN/dp_T$. This is, however, valid procedure only when $dN/dp_T$ is
of exponential form. Therefore any deviation from such behaviour 
are always under detailed scrutiny in which one is searching for the
possible causes. In \cite{ALQ} it was suggested that the extreme
conditions of high density and temperature occuring in
ultrarelativistic heavy ion collisions can lead to memory effects 
and long-range colour interactions and to the presence of
non-Markovian processes in the corresponding kinetic equations
\cite{EXAMPLES}. Actually, as has been shown in \cite{ALQ,UWW},         
$dN(p_T)/dp_T$ are best described, cf., Fig. 2, by a slightly
nonexponential function of the type 
\begin{equation}
\frac{dN(p_T)}{dp_T}\, =\, {\rm const}\cdot \left[ 1\, - (1 - q)
\frac{\sqrt{m^2 + p^2_T}}{kT}\right]^{\frac{1}{1-q}}\qquad
\stackrel{q\rightarrow 1}{\Longrightarrow}\qquad {\rm const}\cdot
\exp\left( - \frac{\sqrt{m^2 + p^2_T}}{kT}\right) . \label{eq:Pt}
\end{equation}
Here $m$ is the mass of produced particle, $k$ is the Boltzmann
constant (which we shall, in what follows, put equal unity) and $T$ is,
for the $q=1$ case, the {\it temperature} of the reaction considered
(or, rather, the temperature of the hadronic system produced). 
The deviation from the exponential form one finds is very small, on
the level $q=1.015$. However, as we shall demonstrate later it can
lead to a quite dramatic effects. It was also shown in \cite{ALQ}
that to first order in $|q-1|$ the generalized slope becomes the
quantity 
\begin{equation}
T_q\, =\, T\, +\, (q-1) m_T . \label{eq:TQ}
\end{equation}
with $T$ being temperature of {\it a purely thermal source}.
This should be contrasted with the empirical relation for the slope
parameter $T$, from which the freeze-out temperature (at which
hadrons are created from the QGP) $T_f$ is then deduced,
\begin{equation}
T\, =\, T_f\, +\, m\langle v_{\perp}\rangle^2 . \label{eq:Tf}
\end{equation}
\begin{figure}[h]
\setlength{\unitlength}{1cm}
\begin{picture}(25.,16.5)
\includegraphics{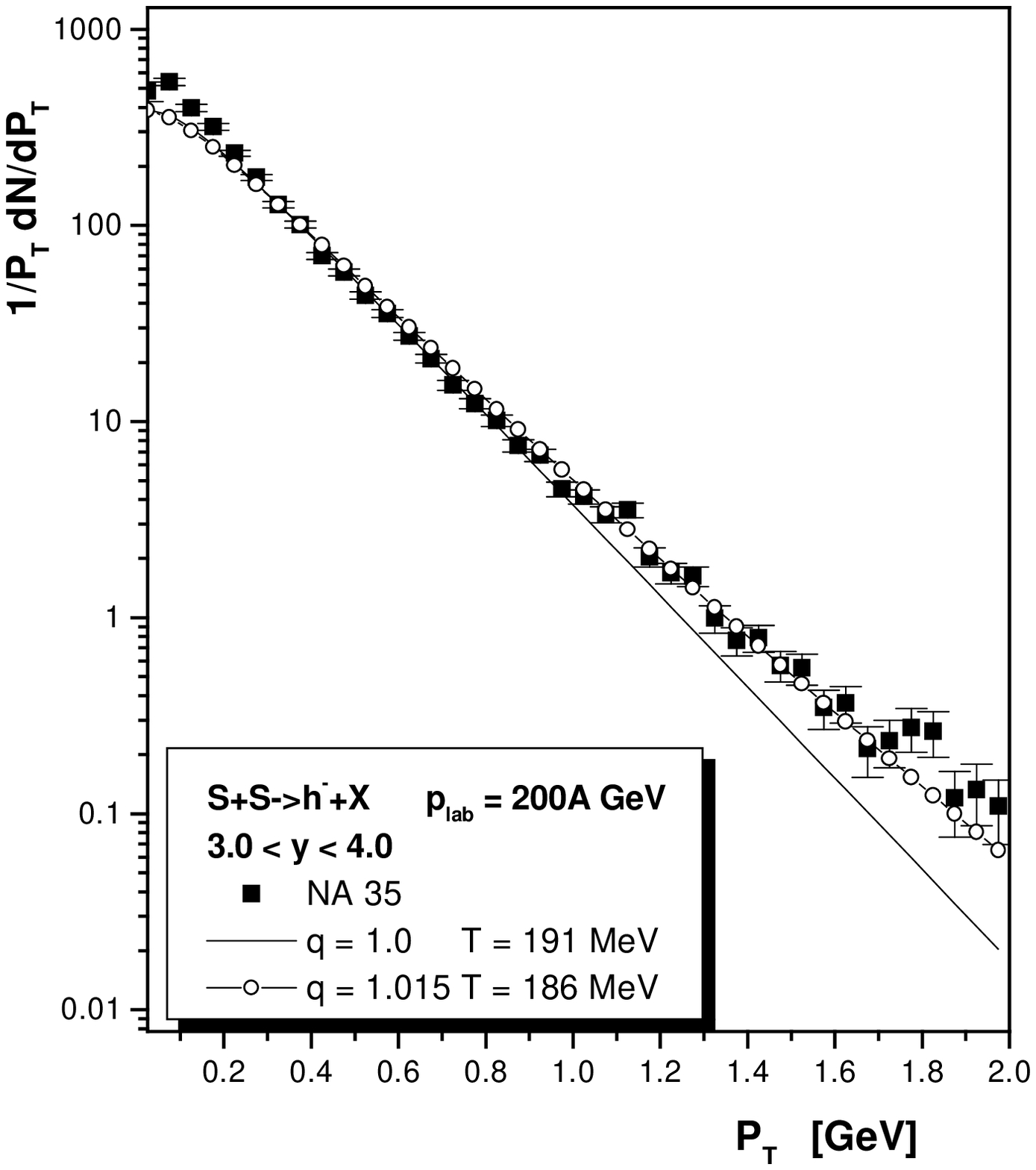}
\end{picture}
\end{figure}
\begin{itemize}
 \item[Fig. 2] The results for $p_T$ distribution 
               $dN(p_T)/dp_T$: notice that
               $q=1.015$ results describes also the tail of
               distribution not fitted by the conventional
               exponent (i.e., $q=1$). This figure is
               reproduced from Fig. 3 of \cite{UWW}.
\end{itemize}
The $\langle v_{\perp}\rangle$ is a fit parameter usually identified with
the average collective (transverse) flow velocity of the hadrons
being produced. In (\ref{eq:TQ}) one has, instead, {\it a purely
thermal source} experiencing a kind of blue shift at high $m_T$
(actually increasing with $m_T$). The nonextensivity parameter $q$
accounts here for all possibilities one can find in \cite{EXAMPLES}
and could therefore be regarded as a new way of presenting
experimental results with $q\neq 1$, signaling that there is something
going on in the collision that prevents it from being exactly
thermal-like in the ordinary sense mentioned above.\\

To the same cathegory belongs also analysis of the equilibrium
distribution of heavy quarks in Fokker-Planck dynamics expected in
heavy ion collisions \cite{JR}. Not going into details in this case,
we shall only mention that it was found that thermalization of
charmed quarks in a QGP surroundings proceeding via collisions with
light quarks and gluons results in a spectral shape, which can be
described only in $q$ statistics with q=1.114 (for $m_C=1.5$ GeV and
temperature of thermal gluons $T_g=500$ GeV).

\section{Fluctuations as a cause of nonextensivity?}

Guided by the cosmic ray example we would like to discuss now the
hypothesis that the examples of nonextensivity in which exponential
distribution becomes a power-like (L\'evy type) distributions is
caused by the fluctuations in the system which were not accounted for
and which make $q \neq 1$ \cite{WW,DENTON}. In other words, $q$
summarizes the action of averaging over $1/\lambda$:
\begin{equation}
\left\langle \exp\left[ - \left(\frac{1}{\lambda}\right) \varepsilon
    \right]\right\rangle \Longrightarrow \left[ 1 \, -\, (1 -
q)\frac{1}{\lambda_0}\varepsilon \right]^{\frac{1}{1-q}} , \label{eq:average}
\end{equation}
where $1/\lambda_0 = \langle 1/\lambda\rangle$.  This connection,
which is clear in the cosmic ray example, should also be true for the
heavy ion collision. Actually, this later case is even more important
and interesting because of the long and still vivid discussion on the
possible dynamics of temperature fluctuations \cite{TFLUC} and
because of its connection with the problem of QGP production in heavy
ion collisions \cite{FLUCTQ}. 
  
Actually the conjecture (\ref{eq:average}) is known in the literature
as Hilhorst integral formula \cite{C} but was never used in the
present context. It says that (for $q>1$ case, where $\varepsilon \in
(0, \infty)$)
\begin{equation}
\left( 1\, +\, \frac{\varepsilon}{\lambda_0}\, \frac{1}{\alpha}\right)^{-a}\, =\,
\frac{1}{\Gamma(\alpha)}\, \int^{\infty}_0\, d\eta\, 
      {\eta}^{\alpha - 1}\, \exp\left[ - \eta\, 
      \left(1\, +\, \frac{\varepsilon}{\lambda_0}\,
       \frac{1}{\alpha}\right)\right].
 \label{eq:GF}
\end{equation}
where $\alpha = \frac{1}{q-1}$. Changing variables under the integral
to $\eta=\alpha \frac{\lambda_0}{\lambda}$, one obtains 
\begin{equation}
L_{q>1}(\varepsilon;\lambda_0)\, =\, C_q\, 
\left( 1\, +\, \frac{\varepsilon}{\lambda_0}\, \frac{1}{\alpha}\right)^{-a}\, =\,
C_q\, \int^{\infty}_0\, \exp\left( - \frac{\varepsilon}{\lambda}\right)\,
f\left(\frac{1}{\lambda}\right)\, d\left(\frac{1}{\lambda}\right)
\label{eq:DEF} 
\end{equation}
where $f(1/\lambda)$ is given by the following gamma distribution:
\begin{equation}
f_{q>1}\left(\frac{1}{\lambda}\right)\, =\,
f_{\alpha}\left(\frac{1}{\lambda},\frac{1}{\lambda_0}\right)\, =\,
\frac{\mu}{\Gamma(\alpha)}\,
\left(\frac{\mu}{\lambda}\right)^{\alpha-1}\, \exp\left(
- \frac{\mu}{\lambda}\right) \label{eq:F}
\end{equation}
with $\mu = \alpha \lambda_0$ and with mean value and variation in the form:
\begin{equation}
\left\langle \frac{1}{\lambda}\right\rangle \, =\,
 \frac{1}{\lambda_0} \qquad {\rm and}\qquad
\left\langle \left(\frac{1}{\lambda}\right)^2\right\rangle\, -\, 
\left\langle\frac{1}{\lambda}\right\rangle^2\, =\, 
\frac{1}{\alpha\, \lambda_0^2} . \label{eq:MEANVAR}
\end{equation}
Notice that, with increasing $\alpha$ the variance (\ref{eq:MEANVAR})
decreases and asymptotically (for $\alpha \rightarrow \infty$, i.e,
for $q\rightarrow 1$) the gamma distribution (\ref{eq:F}) becomes
a delta function, $f_{q>1}(1/\lambda)=\delta (\lambda - \lambda_0)$. The
relative variance for this distribution is given by 
\begin{equation}
\omega\, =\, \frac{\left\langle\left(\frac{1}{\lambda}\right)^2\right\rangle\, 
 -\, \left\langle\frac{1}{\lambda}\right\rangle^2}
 {\left\langle \frac{1}{\lambda}\right\rangle^2}\, =\,
\frac{1}{\alpha}\, =\, q\, -\, 1 . \label{eq:PROOF}
\end{equation}

For the $q<1$ case $\varepsilon$ is limited to $\varepsilon \in
[0,\lambda_0/(1-q)]$. Proceeding in the same way as before, i.e.,
making use of the following representation of the Euler gamma function
(where $\alpha' = - \alpha = \frac{1}{1-q}$)
\begin{equation}
\left[1\, -\, \frac{\varepsilon}{\alpha' \lambda_0}\right]^{\alpha'}\, =\,
\left(\frac{\alpha' \lambda_0}{\alpha' \lambda_0 -
\varepsilon}\right)^{-\alpha'}\, 
=\, \frac{1}{\Gamma(\alpha')}\, \int^{\infty}_0\, d\eta\, 
    \eta^{\alpha' - 1}\, \exp\left[ - \eta\, \left(1\, +\, 
    \frac{\varepsilon}{\alpha' \lambda_0 - \varepsilon}\right)\right]
      , \label{eq:EGF}
\end{equation}
and changing variables under the integral to $\eta = \frac{\alpha'
\lambda_0 - \varepsilon}{\lambda}$, we obtain
$L_{q<1}(\varepsilon;\lambda_0)$ in the form of eq. (\ref{eq:DEF}) but
with $\alpha \rightarrow -\alpha'$ and with the respective $f(1/\lambda)
= f_{q<1}(1/\lambda)$ given now by the same gamma distribution as in
(\ref{eq:F}) but this time with $\alpha \rightarrow \alpha'$ and
$\mu = \mu(\varepsilon) = \alpha' \lambda_0 - \varepsilon$. Contrary to the
$q>1$ case, this time the fluctuations depend on the value of the
variable in question, i.e., the mean value and variance are now both
$\varepsilon$-dependent:  
\begin{equation}
\left\langle \frac{1}{\lambda}\right\rangle\, =\, \frac{1}{\lambda_0 -
\frac{\varepsilon}{\alpha'}}\qquad {\rm and}\qquad \left\langle
\left(\frac{1}{\lambda}\right)^2\right\rangle\, -\,
\left\langle\frac{1}{\lambda}\right\rangle^2\, =\, \frac{1}{\alpha'}\cdot
\frac{1}{\left(\lambda_0 - \frac{\varepsilon}{\alpha'}\right)^2} .
\label{eq:MV}
\end{equation}
However, the relative variance
\begin{equation}
\omega\, =\,  \frac{\left\langle\left(\frac{1}{\lambda}\right)^2\right\rangle\,
       -\, \left\langle\frac{1}{\lambda}\right\rangle^2}
       {\left\langle \frac{1}{\lambda}\right\rangle^2}\, 
       =\, \frac{1}{\alpha'}\, =\, 1\, -\, q , \label{eq:RESULT}
\end{equation}
remains $\varepsilon$-independent and depends only on the parameter
$q$. As above the resulting gamma distribution becomes a delta
function, $f_{q<1}(1/\lambda)=\delta (\lambda - \lambda_0)$, for
$\alpha' \rightarrow \infty$, i.e., for $q\rightarrow 1$.\\  

Summarizing: one can indeed say that the nonextensivity parameter $q$
in the $L_q(\varepsilon)$ distributions can be expressed by
the relative variance $\omega$ of fluctuations of the parameter
$1/\lambda$ in the distribution $L_{q=1}(\varepsilon)$: 
\begin{equation}
q = 1 \, \pm \, \omega \qquad {\rm for}\quad q>1~(+)\quad {\rm and}
\quad q<1~(-) .       \label{eq:QG1}
\end{equation}

Actually the above result (\ref{eq:QG1}) is derived for the particular
(Gamma-like) shape of the fluctuation distribution function. How can
it be realized? To answer this question let us notice
\cite{WW,DENTON} that in the case when stochastic variable $\lambda$
is decribed by the usual Langevin equation 
\begin{equation}
\frac{d\lambda}{dt}\, +\, \left[\frac{1}{\tau}\, +\, \xi(t)\right]\,
\lambda\, =\, \phi\, =\, {\rm const}\, >\, 0 , \label{eq:LE}
\end{equation}
(with damping constant $\tau$ and source term $\phi_{q<1}\, =\,
\frac{1}{\tau}\left(\lambda_0 - \frac{\varepsilon}{\alpha'}\right)$ or
$\phi = \phi_{q>1} = \frac{\lambda_0}{\tau}$), then for the stochastic
processes defined by the {\it white gaussian noise} form of 
$\xi(t)$\footnote{It means that ensemble mean $\langle \xi(t)
\rangle\, =\, 0 $ and correlator (for sufficiently fast changes)
$\langle \xi(t)\, \xi(t + \Delta t) \rangle\, =\, 2\, D\,
\delta(\Delta t)$. Constants $\tau$ and $D$ define, respectively, the
mean time for changes and their variance by means of the following
conditions: $\langle \lambda(t)\rangle\, =\, \lambda_0\, \exp\left( -
\frac{t}{\tau} \right)$ and $\langle \lambda^2(t=\infty)\rangle\, =\,
\frac{1}{2}\, D\, \tau$. Thermodynamical equilibrium is assumed here
(i.e., $t >> \tau$, in which case the influence of the initial condition
$\lambda_0$ vanishes and the mean squared of $\lambda$ has value
corresponding to the state of equilibrium).} one obtains the
following Fokker-Plank equation \cite{FP} for the distribution
function of the variable $\lambda$:
\begin{equation}
\frac{df(\lambda)}{dt}\, =\, -\, \frac{\partial}{\partial \lambda}K_1\,
f(\lambda)\, +\, \frac{1}{2}\, \frac{\partial^2}{\partial \lambda^2}K_2\,
f(\lambda) , \label{eq:FPE}
\end{equation}
with intensity coefficients $K_{1,2}$ defined by eq.(\ref{eq:LE}) and
equal to (cf., \cite{ADT}):  
\begin{equation}
K_1(\lambda)\, =\, \phi\, -\, \frac{\lambda}{\tau}\, +\, D\, \lambda
\qquad {\rm and}\qquad 
K_2(\lambda)\, =\, 2\, D\, \lambda^2 . \label{eq:KK}
\end{equation}
Its stationary solution has precisely the Gamma-like form we are looking for:
\begin{equation}
f(\lambda)\, =\, \frac{c}{K_2(\lambda)}\, \exp\left[\, 2\,
\int^{\lambda}_0 d\lambda'\, \frac{K_1(\lambda')}{K_2(\lambda')}\, \right]
\, =\, \frac{1}{\Gamma(\alpha)}\, \mu\, 
 \left(\frac{\mu}{\lambda}\right)^{\alpha-1}\, \exp\left( -\,
\frac{\mu}{\lambda} \right) , \label{eq:FRES}
\end{equation}
with the constant $c$ defined by the normalization condition,
$\int^{\infty}_0 d(1/\lambda) f(1/\lambda) = 1$. It depends on two
parameters:  
\begin{equation}
\mu(\varepsilon)\, =\, \frac{\phi_q(\varepsilon)}{D} \qquad {\rm and}
\qquad \alpha_q\, =\, \frac{1}{\tau\, D} ,\label{eq:PAR}
\end{equation}
with $\phi_q = \phi_{q>1,q<1}$ and $\alpha_q = (\alpha, \alpha')$
for, respectively, $q>1$ and $q<1$. Therefore eq. (\ref{eq:QG1}) with
$\omega = \frac{1}{\tau D}$ is, indeed, a sound possibility with (in
the case discussed above) parameter of nonextensivity $q$ given
by the parameter $D$ and by the damping constant $\tau$ describing
the {\it white noise}.\\    

\section{Back to heavy ion collisions}

With the possibility of such interpretation of parameter $q$ in
mind we can now came back to the example of heavy ion collisions and
see what are its consequences in this case. It is interesting to notice
that the relatively small departure of $q$ from unity, $q-1 \simeq
0.015$ \cite{ALQ,UWW}, if interpreted in terms of the
previous section, indicates that rather large relative fluctuations of
temperature, of the order of $\Delta T/T \simeq 0.12$, exist in
nuclear collisions. It could mean therefore that we are dealing here
with some fluctuations existing in small parts of the system in
respect to the whole system (according to interpretation of \cite{L})
rather than with fluctuations of the event-by-event type in which,
for large multiplicity $N$, fluctuations $\Delta T/T = 0.06/
\sqrt{N}$ should be negligibly small \cite{FLUCTQ}. This controversy
could be, in principle, settled by detailed analyses of the
event-by-event type. Already at present energies and nuclear targets
(and the more so at the new accelerators for heavy ions like RHIC at
Brookhaven, now commisioned, and LHC at CERN scheduled to be
operational in the year 2006) one should be able to check whether the
power-like $p_T$ distribution $dN(p_T)/dp_T$ occurs already at every
event or only after averaging over all events. In the former case we
would have a clear signal of thermal fluctuations of the type
mentioned above. In the latter case one would have for each event a
fixed $T$ value which would fluctuate from one event to another (most
probably because different initial conditions are encountered in a given
event).\\  

One point must be clarified, however. The above conjecture rests on
the stochastic equation (\ref{eq:LE}). Can one expect such equation
to govern the $T$ fluctuations? To ansewer this question let us turn
once more to the fluctuations of temperature \cite{TFLUC,FLUCTQ,L}
discussed before, i.e., to $\lambda = T$. Suppose that we have a
thermodynamic system, in a small (mentally separated) part of which
the temperature fluctuates with $\Delta T \sim T$. Let $\lambda(t)$
describe stochastic changes of the temperature in time. If the mean
temperature of the system is $\langle T\rangle = T_0$ then, as result
of fluctuations in some small selected region, the actual temperature
equals $T' = T_0 - \tau \xi(t) T$. The inevitable exchange of heat
between this selected region and the rest of the system leads to the
equilibration of the temperature and this process is described by the
following equation \cite{LLH}:
\begin{equation}
\frac{\partial T}{\partial t}\, -\, \frac{1}{\tau}
            \, (T'\, -\, T)\, +\, \Omega_q = 0 ,\label{eq:HC}
\end{equation}
which is, indeed, of the type of eq. (\ref{eq:LE}) (here
$\Omega_{q<1} = \frac{\varepsilon}{\tau \alpha'}$ and $\Omega_{q>1} 
= 0$). This proves the plausability of what was said above and makes
the event-by-even measurements of such phenomenon very interesting
one.\\ 

\section{Other imprints of nonextensivity}

The above discussed examples do not exhaust the list of the possible
imprints of nonextensivity in multiparticle production know (or
thought of) at present. As it is impossible to review all of them
here, we shall therefore only mention them.\\

In \cite{ALQ,UWW} the possible effects of nonextensivity on the mean
occupation numbers $n_q$ and on the event-by-event fluctuation
phenomena have been discussed. It turns out, for example, that some
charateristics of fluctuations are extremaly sensitive to even small
departures of $q$ from unity. Such departure can easily mimick the
existing correlations or the influence of resonances.\\

In \cite{BCM} an interesting attempt was presented to fit the energy
spectra in both the longitudinal and transverse momenta of particles
produced in the $e^+e^-$ annihilation processes at high energies
using $q$-statistical model. In this way one can have energy
independent temperature $T$ and nonextensivity parameter $q$ rising
quickly with energy from $q=1$ to $q=1.2$ and reflecting long range
correlations in the phase space arising in the hadronization process
in which quarks and gluons combine together forming hadrons
(actually, this observation has general validity and applies to all
production processes discussed here as well). In similar spirit is
the work \cite{Beck} which attemts to generalized the so called
Hagedorn model of multiparticle production to $q$-statistics.\\

To the extend to which self-organized criticality (SOC) is connected
with nonextensivity \cite{T} one should also mention here a very
innovative (from the point of view of high energy collision)
application of the concept of SOC to such processes \cite{MENG}. \\

The other two examples do not refer to Tsallis thermostatistics
directly, nevertheless they are connected to it. First is the attempt
to study, by using the formalism of quantum groups, the Bose-Einstein
correlations between identical particles observed in multiparticle
reactions \cite{BECQ}, second are works on intermittency and
multiparticle distributions using the so called L\'evy stable
distributions \cite{INTER}. They belong, in some sense, to the domain
of nonextensivity because, as was shown in \cite{TQ}, there is 
close correspondence between the deformation parameter of quantum
groups used in \cite{BECQ} and the nonextensivity parameter $q$ of
Tsallis statistics and there is also connection between Tsallis
statistics and L\'evy stable distributions \cite{INTER}. Some traces of
the possible nonextensive evolution of cascade type hadronization
processes were also searched for in \cite{UWWC}. The quantum group
approach \cite{BECQ,TQ} could probably be a useful tool when studying 
delicate problem of interplay between QGP and hadrons produced from
it. It is plausible that description in terms of $q$-deformed bosons
(or the use of some kind of interpolating statistics) would lead to
more general results than the  simple use of nonextensive mean
occupation numbers $<n>_q$ discussed above (for which the only known
practical description is limited to small deviations from
nonextensivity only). \\ 

One should mention also attempts to use L\'evy-type distributions to
fit the development of the cosmic ray cascades to learn of how many
descendat they contain \cite{RWW} or the explanation of the Feynman
scaling violation observed in multiparticle distributions in terms of
the $q$ parameter \cite{UWWF}. \\

\section{Summary}

To summarize, evidence is growing in favour of the view that the
standart statistical model can be enlarged towards the nonstandart
statistics and that by including one new parameter $q$ it allows to
repoduce much broader set of data than it was done so far. It was
demonstrated that $q$ is probably connected with the intrinsic
fluctuations existing in the system under consideration which were
previously not considered at all\footnote{Actually it is likely that
the importance of fluctuations is more general than their
applications to $q$-statistics only. A good example is work
\cite{Bialas} showing how fluctuations of the vacuum can change the
theoretically expected gaussian spectra in string models into the
observed exponential one with a kind of effective "temperature".}.
Let us close with the remarks that, as was shown in \cite{QFT}, one can
also try to use power-like distributions discussed here to the new
formulation of the quantum field theory (for example, in terms of
lorenzian rather than gaussian path integrals, accounting in this
manner for some intrinsic long range correlations impossible to
dispose of and extremaly difficult to account for by other methods).\\

Acknowledgements: One of us (G.W.) would like to thank the organizers
of the RANP2000 for support making his participation in this workshop
possible. Authors are grateful to Prof. S.Abe for his valuable
comments.\\


\noindent
{\bf Discussion}\\

\noindent
Question by K.Werner: {\it Concerning your example $dN/dp_T$:
this seems to be quite under control,large $p_T$ is perturbative QCD
(power-like), small $p_t$ is soft, so you have a superposition of
these two processes. There seems to be no need of these additional
fluctuations.} \\
Answer: Your proposition is just one of many around us. You are right
that adding two known (in their domain of applicability) mechanisms
will also fit data. But what we are saying here is that, perhaps, there
is mechanism which would show itself in the intermediate region and
which can be described by the nonextensive parameter $q$. And it can
be experimentally checked in event-by-event type of analysis, as was
presented above. This would answer the question is there a need for
something or not. Besides, it should be stressed that using 
only one new parameter $q$ one can fit quite large (in \cite{BCM} 
the whole) region of $p_T$.\\

\noindent
Question by F.Grassi: {\it Did you try to fit $p_T$ distributions
for other type of particles than charged?}\\
Answer: No, at least not yet.\\

\noindent
Question by T.Kodama: {\it If $q$ represents the measure of
fluctuations of ensemble, then it would also reflect in multiplicity
distributions. However, value of $q$ is different for the $p_T$
distribution and multiplicity distribution. How do you interpret
this?} \\
Answer: It is difficult for me to comment because so far I have not
seen such analysis of the multiplicity data. But if things are really
as you say then, assuming that everything was done correctly, I would
argue that multiplicity is global characteristic whereas $p_T$ is a
local one. Therefore they are sensitive to different aspects of the
underlying dynamics and resulting parameters $q$ could {\it a priori} 
be different. One can also say that for $p_T$ distribution the 
situation is more clear as we are simple replacing here $\exp(...)$ 
by $\exp_q(...)$. Multiplicity distributions depend on $q$ in 
indirect way only.\\

\end{document}